\documentstyle[aps,multicol,rotate,epsf]{revtex}
\begin{document}
\draft

\title{Inertial Effects on Fluid Flow through Disordered Porous Media}

\author{U. M. S. Costa$^{1}$, J. S. Andrade Jr.$^{1,2}$, H. A. Makse$^{2}$, and
H. E. Stanley$^{3}$}

\address{$^1$Departamento de F\'{\i}sica, Universidade Federal do Cear\'a,
60451-970 Fortaleza, Cear\'a, Brazil \\
$^2$ Schlumberger-Doll Research, Old Quarry Road, Ridgefield, CT 06877 \\
$^3$ Center for Polymer Studies and Physics Dept., Boston University,
Boston, MA 02215}

\maketitle
\begin{abstract}

We study the fluid flow through disordered porous media by numerically
solving the complete set of the Navier-Stokes equations in a two dimensional 
lattice  with a spatially random distribution of solid obstacles (plaquettes).
We simulate viscous and non-viscous flow through these idealized pore spaces
to determine the origin of the deviations from the classical Darcy's law behavior.
Due to the non-linear contribution of inertia to the transport
of momentum at the pore scale, we observe a typical departure from Darcy's
law at sufficiently high Reynolds numbers. Moreover, we show that the classical 
Forchheimer equation provides a valid phenomenological model to correlate the 
variations of the friction factor of the porous media over a wide range
of Reynolds conditions. 

\end{abstract}

\begin{multicols}{2}
\narrowtext

The application of Darcy's law is the standard approach to characterize
single phase fluid flow in microscopically disordered and macroscopically 
homogeneous porous media \cite{Dul79,Adl92,Sah94}. Basically, one simply assumes 
that a global index, the permeability $k$, relates the average fluid velocity $V$
through the pores, with the pressure drop $\Delta P$ measured across the system,
\begin{equation}
V = -{k \over \mu}{\Delta P \over L}~~,
\label{eq1}
\end{equation}
where $L$ is the length of the sample in the flow direction and $\mu$ is the viscosity 
of the fluid. However, in order to understand the interplay between porous 
structure and fluid flow, it is necessary to examine local aspects of the 
pore space morphology and relate them with the relevant mechanisms of momentum 
transfer (viscous and inertial forces). This has been accomplished in previous 
studies \cite{Sch93,Mar94} where computational simulations based on the knowledge of the pore 
space morphology have been quite successful in predicting permeability 
coefficients of real porous materials.

In spite of its great applicability, the concept of permeability 
as a global index for flow, which implies the validity of Eq.~(1), should 
be restricted to viscous flow conditions or, more precisely, to low values 
of the Reynolds number, $Re \equiv \rho V d_p /\mu$, where $\rho$
is the density of the fluid and $d_p$ is the grain 
diameter. Unlike the sudden transition from laminar to turbulent flow in pipes 
and channels, where there is a critical Reynolds number condition separating 
both regimes, experimental studies have shown that the passage from the linear 
(Darcy's law) to the nonlinear behavior in flow through porous media is more likely 
to be gradual (see Dullien \cite{Dul79} and references therein).
It has then been argued that the contribution of inertial forces 
(convection) to the flow in the pore space should also be examined in the framework 
of the laminar flow regime before assuming that fully developed turbulence effects 
are already present and relevant to momentum transport in the system. Recently \cite{And95}, 
the transport of momentum in two-dimensional porous structures generated at the 
critical percolation point \cite{Sah94} has been evaluated through the direct solution of the 
Navier-Stokes and continuity equations. It has been shown that, beyond the range 
of validity of Darcy's law, a nonuniversal behaviour should be considered for the 
critical exponent $t$ relating flow permeability coefficients and porosities, 
$k \propto {(\epsilon-{\epsilon}_c)}^t$, with ${\epsilon}_c$ being the critical 
percolation porosity. In the present work we show by numerical simulation of 
the Navier-Stokes equations in low porosity structures ($\epsilon>{\epsilon}_c$)
that the departure from Darcy's law at sufficiently high $Re$ 
numbers can be explained in terms of the inertial contribution to {\it laminar fluid 
flow} through the void space. In other words, we demonstrate that there is no need to 
include turbulence effects to model incipient deviations from linearity usually 
found in permeability experiments with various porous materials \cite{Dul79}.

Our topological model for the pore connectivity is based on the general 
picture of site percolation disorder. Square obstacles are randomly removed from a 
64x64 square lattice until a porous space with a prescribed void fraction $\epsilon$ 
is generated. The mathematical description for the detailed fluid mechanics 
in the interstitial pore space is based on the assumptions that we have steady 
state flow in isothermal conditions and the fluid is continuum, Newtonian and 
incompressible. Thus, the essence of our phenomenological description is the 
two-dimensional set of Navier-Stokes and continuity equations for momentum 
and mass conservation,      
$$
\rho\left[u{\partial u \over \partial x} + v {\partial u\over 
\partial y}
\right] = -{\partial P\over \partial x} 
+\mu\left[{\partial^2u\over\partial 
x^2} + {\partial^2u\over \partial y^2}\right]~~,
$$ 
$$
\rho\left[u{\partial v \over \partial x} + v {\partial v \over 
\partial y}
\right] = -{\partial P\over \partial y} +\mu\left[{\partial^2v 
\over\partial 
x^2} + {\partial^2v \over \partial y^2}\right]~~,
$$
\begin{equation}
{\partial u\over \partial x} + {\partial v \over \partial y} = 
0.
\label{eq2}
\end{equation}
Here $u$ and $v$ are the components of the velocity vector in the
$x$ and $y$ directions, respectively. We use the nonslip boundary condition
at the whole of the solid-fluid interface. 
End effects of the flow field established inside the pore structure (particularly 
significant at high $Re$ conditions) are minimized by attaching a header (inlet) and 
a recovery (outlet) region to two opposite faces arbitrarily chosen. At the inlet 
line, a constant inflow velocity in the normal direction to the boundary is 
specified whereas at the exit, the rate of the velocity change is assumed to be 
zero (gradientless boundary condition). Instead of periodic boundary conditions, 
we choose to close the remaining two faces of the system with two additional 
columns of obstacles. This insulating condition reproduces more closely the 
experimental setup usually adopted with real rocks and permeameters. For a 
given realization of the pore geometry and a fixed $Re$, the velocity and pressure 
fields in the fluid phase of the void space and ancillary zones are 
numerically 
obtained through discretization by means of the control volume finite-difference 
technique \cite{Pat80}. Finally, from the area-averaged pressures at the inlet and outlet 
positions, the overall pressure drop across the porous realizations can be readily calculated.

\begin{figure}
\centerline{
\vbox{ \hbox{\epsfxsize=5.cm
 \rotate[r]{\epsfbox{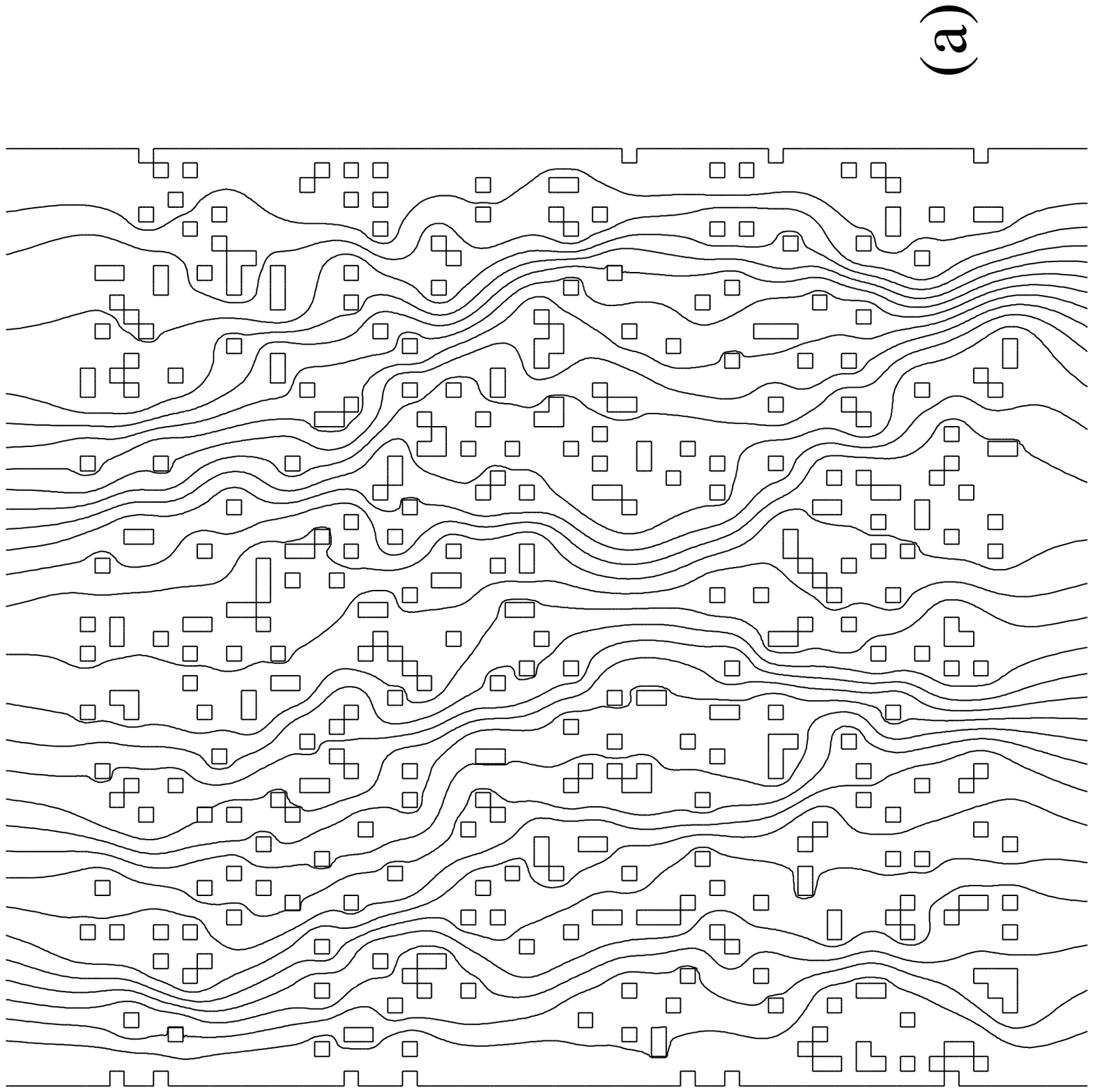}}}
\vspace{.5cm}
\vbox{ \hbox{\epsfxsize=5.cm
 \rotate[r]{\epsfbox{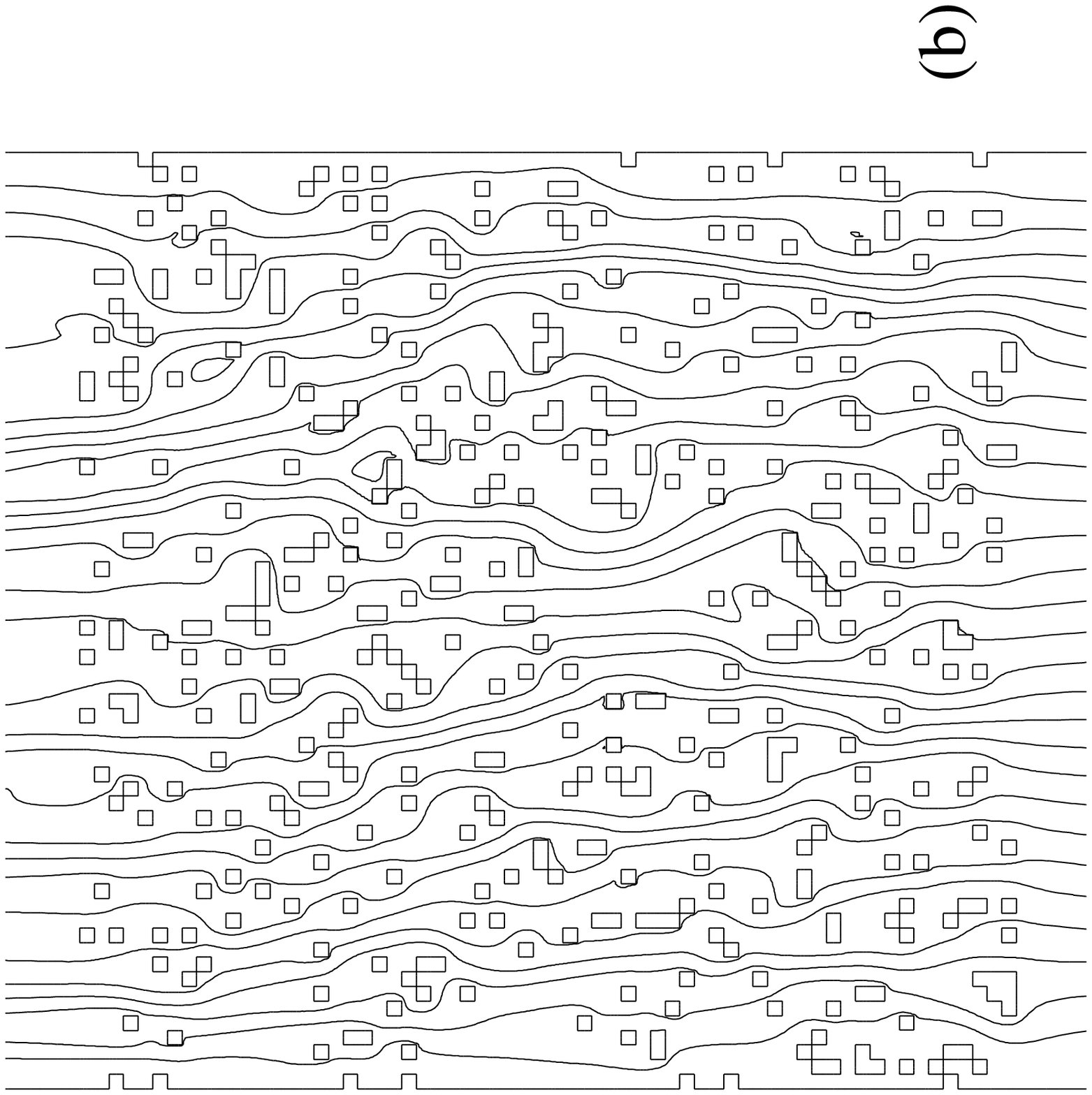}}}
}
}}
\vspace{0.5cm}
\narrowtext
\caption{
(a) Contour plot of the stream function $\psi$ for low Reynolds number 
conditions ($Re=0.0156$).  (b) Same as in (a), but for a higher Reynolds
number ($Re=15.6$).  
}
\label{f.1}
\end{figure}

Figure 1a  shows the contour plot of the stream function $\psi$ \cite{stream} 
for a typical realization of a highly porous void space ($\epsilon=0.9$) subjected to low 
Reynolds conditions, $Re=0.0156$. In spite of the highly connected pathway available for flow 
at this large porosity value, the predominant viscous forces in the momentum transport 
through the complex void geometry generates well defined ``preferential channels'' of fluid 
flow. As shown in Fig.~1b, the situation is quite different at high Reynolds conditions, 
where the degree of channeling is clearly less intense than in Fig.~1a. 
In the case of Fig.~1b, due to the relevant contribution of inertial forces (convection) to the 
flow at the pore scale, the distribution of streamlines along the direction orthogonal 
to the main flux becomes more homogeneous. This nonlinear effect can be macroscopically 
quantified in terms of the so-called Forchheimer equation \cite{Dul79,Sah94},
\begin{equation}
-{\Delta P \over L} = {\alpha \mu V} + {\beta \rho {V^2}}~~,
\label{eq3}
\end{equation}
where the coefficient $\alpha$ should correspond to the reciprocal permeability of the 
porous material and $\beta$ is usually named as an ``inertial parameter''. Equation (3) with 
constant $\alpha$ and $\beta$ parameters is not a purely empirical expression since it can 
be derived by an adequate average of the Navier-Stokes equation for steady and incompressible 
laminar flow of a Newtonian fluid in a rigid porous medium \cite{Dul79}. Rearranging the 
Forchheimer equation in the form,
\begin{equation}
f = {1 \over Re'} + 1~~,
\label{eq4}
\end{equation}
where $f\equiv-\Delta P / {L \beta \rho {V^2}}$ and $Re'\equiv\beta \rho V / {\alpha \mu}$,
we obtain a {\it friction factor-Reynolds number}
type of correlation which is presumably ``universal''. Indeed, Eq.~(4) has been 
extensively and successfully used to correlate experimental data from a large 
variety of porous materials and a broad  range of flow conditions \cite{Dul79}. 
In Fig.~2, we show the results of simulations performed with three realizations
of the porous structure generated with $\epsilon=0.9$.
After computing and averaging the overall pressure drops for all realizations
at different $Re$ numbers, we estimate the coefficients $\alpha$ and $\beta$
and calculate the modified variables $f$ and $Re'$. 
In agreement with real flow experiments, Eq.~(4) also provides a 
satisfactory fit to all results of our computational simulations. Moreover, 
the point of departure from linear (Darcy's law) to nonlinear behavior 
in the range ${10^{-2}}<Re'<{10^{-1}}$ of modified Reynolds is also consistent
with previous experimental observations.

In an attempt to characterize the influence of inertial forces on 
the flow behavior of a single fluid in highly porous structures, we demonstrate 
here, by direct simulation of the Navier-Stokes equations, that incipient 
deviations from Darcy's law observed in several experiments, can be 
satisfactorily modeled in the laminar regime of fluid flow, without
including turbulence effects. The results of our computational simulations 
corroborates numerous experimental data which display a gradual 
transition at high $Re$ from linear to nonlinear flow in the 
pore space. In addition, we show that the physical 
description underlying the classical Forchheimer equation provides a legitimate 
correlation for the global friction factor of the porous media over a wide range
of Reynolds number conditions. In summary, whether true turbulence effects have 
been detected or not in real flows through porous media, our calculations with 
the Navier-Stokes equations indicate that the Forchheimer model with constant 
$\alpha$ and $\beta$ parameters remains valid for rather high $Re$ numbers, even 
when convective nonlinearities can significantly affect the momentum transport at 
the pore scale, in comparison to viscous forces. These facts might be relevant to 
understand the role played by convection in several technical applications involving 
consolidated and non-consolidated porous materials. 

\bigskip

\noindent
We thank CNPq and FUNCAP for support.

\begin{figure}
\centerline{
\vbox{ \hbox{\epsfxsize=5.cm
 \rotate[r]{\epsfbox{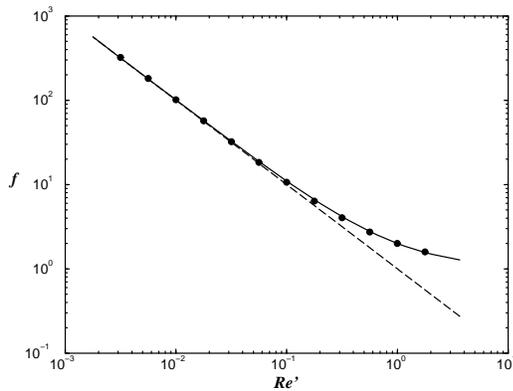}}}
}
}
\vspace{0.5cm}
\narrowtext
\caption{
Dependence of the generalized  friction factor $f$ on the modified 
Reynolds number $Re'$. The solid line is the best fit to the 
data of the Forchheimer equation. The dashed line is the best
fit to the data at low $Re'$ of Darcy's law. 
}
\label{f.2}
\end{figure}

\end{multicols}


\begin{references}
\bibitem{stream} The stream function $\psi$ is defined for incompressible
two-dimensional flows as $u\equiv{\partial\psi/\partial y}$ and 
$v\equiv-{\partial\psi/\partial x}$.
\end{references}
\end{document}